\documentclass[%
reprint, 
superscriptaddress,
nofootinbib,
amsmath,amssymb,
aps,
prl
]{revtex4-2}

\usepackage{graphicx}
\usepackage{dcolumn}
\usepackage{bm}
\usepackage{xcolor}
\usepackage{hyperref}
\usepackage{chemmacros}
\usepackage{todonotes}

\begin{document}


\title{Lattice-charge coupling in a trilayer nickelate with intertwined density wave order}
\author{Xun Jia}
\email{jiaxun@ihep.ac.cn}
\affiliation{Materials Science Division, Argonne National Laboratory, Lemont, IL 60439, USA
}
\affiliation{Multi-disciplinary Research Division, Institute of High Energy Physics, Chinese Academy of Sciences, Beijing 100049, China}

\author{Yao Shen}
\altaffiliation{Present address: Institute of Physics, Chinese Academy of Sciences, Beijing 100190, China}
\affiliation{
 Condensed Matter Physics and Materials Science Department, Brookhaven National Laboratory, Upton, NY 11973, USA
}

\author{Harrison LaBollita}
\altaffiliation{Present address: Center for Computational Quantum Physics, Flatiron Institute, New York, NY 10010, USA}
\affiliation{
 Department of Physics and Astronomy, Arizona State University, Tempe, AZ 85218, USA
}%

\author{Xinglong Chen}
\altaffiliation{Present address: School of Physics, Southeast University, Nanjing, Jiangsu 210096, China}
\affiliation{Materials Science Division, Argonne National Laboratory, Lemont, IL 60439, USA
}

\author{Junjie Zhang}
\altaffiliation{Institute of Crystal Materials, State Key Laboratory of Crystal Materials, Shandong University, Jinan, Shandong 250100, China}
\affiliation{Materials Science Division, Argonne National Laboratory, Lemont, IL 60439, USA
}

\author{Yu Li}
\affiliation{Materials Science Division, Argonne National Laboratory, Lemont, IL 60439, USA
}

\author{Hengdi Zhao}
\affiliation{Materials Science Division, Argonne National Laboratory, Lemont, IL 60439, USA
}

\author{Mercouri G. Kanatzidis}
\affiliation{Materials Science Division, Argonne National Laboratory, Lemont, IL 60439, USA}
\affiliation{Department of Chemistry, Northwestern University, Evanston, IL 60208, USA}

\author{Matthew Krogstad}
\affiliation{%
 X-ray Science Division, Argonne National Laboratory, Lemont, IL 60439, USA
}%

\author{Hong Zheng}
\affiliation{Materials Science Division, Argonne National Laboratory, Lemont, IL 60439, USA
}

\author{Ayman Said}
\affiliation{%
 X-ray Science Division, Argonne National Laboratory, Lemont, IL 60439, USA
}%

\author{Ahmet Alatas}
\affiliation{%
 X-ray Science Division, Argonne National Laboratory, Lemont, IL 60439, USA
}%

\author{Stephan Rosenkranz}
\affiliation{%
 Materials Science Division, Argonne National Laboratory, Lemont, IL 60439, USA
}%

\author{Daniel Phelan}
\affiliation{Materials Science Division, Argonne National Laboratory, Lemont, IL 60439, USA
}

\author{Mark P. M. Dean}
\affiliation{
 Condensed Matter Physics and Materials Science Department, Brookhaven National Laboratory, Upton, NY 11973, USA
}
\affiliation{
Department of Physics and Astronomy, The University of Tennessee, Knoxville, Tennessee 37966, USA
}

\author{M. R. Norman}
\affiliation{%
 Materials Science Division, Argonne National Laboratory, Lemont, IL 60439, USA
}%

\author{J. F. Mitchell}
\affiliation{%
 Materials Science Division, Argonne National Laboratory, Lemont, IL 60439, USA
}%

\author{Antia S. Botana}
\email{antia.botana@asu.edu}
\affiliation{
 Department of Physics and Astronomy, Arizona State University, Tempe, AZ 85218, USA
}%

\author{Yue Cao}
\email{yue.cao@anl.gov}
\affiliation{%
 Materials Science Division, Argonne National Laboratory, Lemont, IL 60439, USA
}%


\begin{abstract}
Intertwined charge and spin correlations are ubiquitous in a wide range of transition metal oxides and are often perceived as intimately related to unconventional superconductivity. Theoretically envisioned as driven by strong electronic correlations, the intertwined order is usually found to be strongly coupled to the lattice as signaled by pronounced phonon softening. Recently, both charge/spin density waves (CDW/SDW) and superconductivity have been discovered in several Ruddlesden-Popper (RP) nickelates, in particular trilayer nickelates \ch{RE4Ni3O10} (RE=Pr, La). The nature of the intertwined order and the role of lattice-charge coupling are at the heart of the debate about these materials. Using inelastic X-ray scattering, we mapped the phonon dispersions in \ch{RE4Ni3O10} and found no evidence of phonon softening near the CDW wavevector over a wide temperature range. Calculations of the electronic susceptibility revealed a peak at the observed SDW ordering vector but not at the CDW wavevector. The absence of phonon softening is in sharp contrast to that in canonical oxide materials, notably cuprates. Our experimental and theoretical findings highlight the crucial role of the spin degree of freedom and establish a foundation for understanding the interplay between superconductivity and density-wave transitions in RP nickelate superconductors and beyond.
\end{abstract}

\maketitle

\section{I. Introduction}

Intertwined charge and spin density waves (CDW and SDW) refer to a quantum phase of matter where strong electronic correlations induce cooperation rather than competition between the charge and spin degrees of freedom \cite{fradkin2015colloquium, agterberg2020physics}. In transition metal oxides, such intertwined order often coexists or resides in the vicinity of unconventional superconductivity. While initially proposed theoretically as an electronic phase of matter, the nature of the intertwined order remains enigmatic in transition metal oxides, especially regarding its interaction with the underlying lattice. Experimentally, the onset of intertwined CDW/SDW order is almost invariably accompanied by sizable distortions of the lattice. These lattice deformations manifest as temperature-dependent phonon dispersion anomalies near the wavevector of the charge order, which were observed prominently across different families of the canonical high $T_c$ cuprates \cite{mcqueeney1999anomalous, reznik2006electron, baron2008first, le2014inelastic, miao2018incommensurate}. These pronounced phonon anomalies appear not only in the higher-energy bond stretching modes but also in the lower-energy branches. Despite extensive studies, whether strong electron-phonon coupling is universal or essential in correlated oxide superconductors, cuprates or otherwise, remains elusive.

The discovery of superconductivity in valence-reduced Ruddlesden-Popper (RP) nickelates \cite{li2019superconductivity, pan2022superconductivity} and more recently in the RP nickelates themselves \cite{sun2023signatures, chen2024electronic, zhu2024superconductivity, zhang2023superconductivity, li2024signature, ko2025, li2025ambient, zhang2025bulk} provides new opportunities for understanding the nature of intertwined order. To date, no robust signatures of spin and charge order occur in infinite-layer nickelates \cite{rossi2022broken, krieger2022charge, tam2022charge,parzyck2024absence}, although both are present in reduced trilayer nickelates \cite{zhang16}. Both spin and charge order were discovered in the bilayer RP nickelate \ch{La3Ni2O7} \cite{xie2024, khasanov2025} and diminished under pressure before the emergence of superconductivity. \ch{RE4Ni3O10} (RE=Pr, La) are among the latest additions to the superconducting nickelates under pressure \cite{zhu2024superconductivity, zhang2023superconductivity, zhang2025bulk}, with a number of electronic properties resembling the cuprates \cite{li2024signature}. The presence of intertwined CDW/SDW order in the trilayer RP nickelates was previously elucidated by elastic X-ray and neutron diffraction \cite{zhang2020intertwined, samarakoon2023bootstrapped}. These experimental findings would suggest strong lattice deformations similar to those observed in cuprates.

Here, we show phonon dispersions in a range of energies near the CDW ordering vector in \ch{RE4Ni3O10} (RE=Pr, La) and demonstrate the surprising absence of phonon anomalies or softening over a wide temperature range using inelastic X-ray scattering (IXS). Calculations of the susceptibility based on the paramagnetic electronic structure reveal a nesting instability at the SDW ordering vector but not at the CDW one. Our observations suggest a relatively minor role of lattice deformations in the trilayer nickelates, which sharply contrasts with the observations in cuprates.  This in turn suggests that the superconductivity observed in these RP nickelate materials has a solely electronic origin.

\section{II. Results}
\subsection{Temperature dependence of the intertwined density waves}

\begin{figure*}
\includegraphics[width=6.75 in]{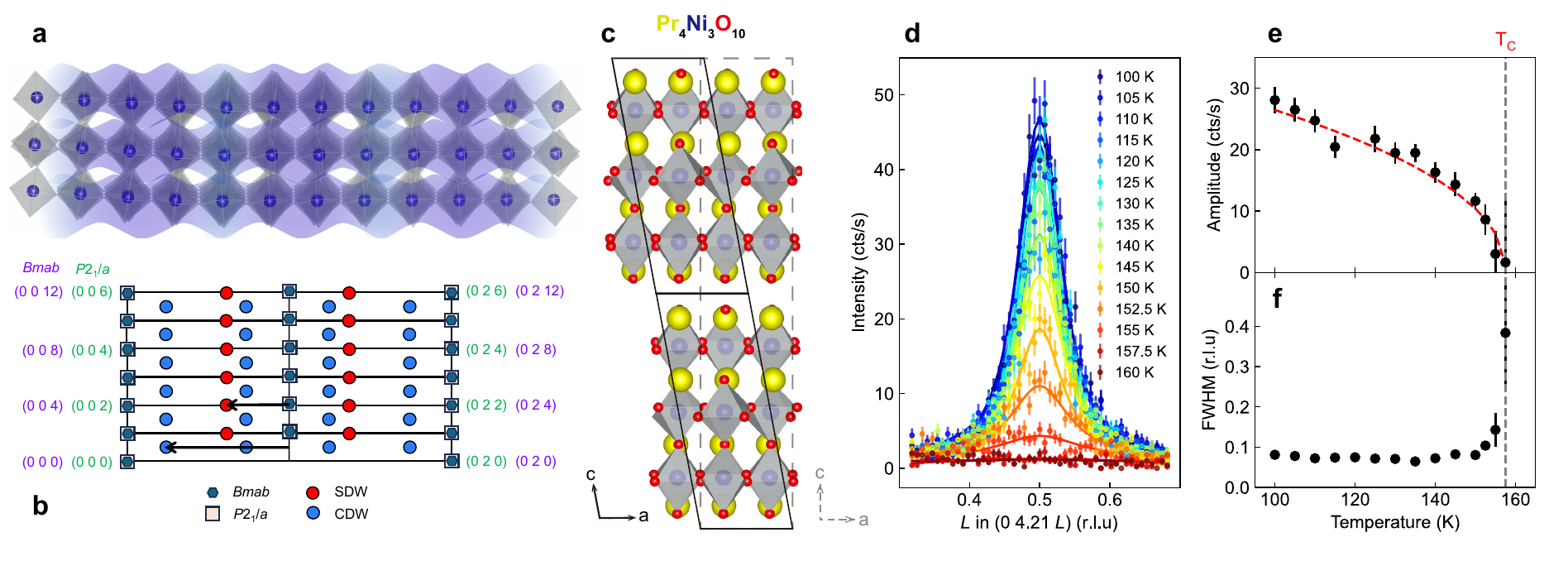} %
\caption{\textbf{Intertwined spin and charge density wave order in Pr$_4$Ni$_3$O$_1$$_0$.} \textbf{a} Illustration of the charge density wave in trilayers of perovskite-like blocks of NiO$_6$. The grey shading illustrates the charge density distribution.
\textbf{b} Schematic of the 0\textit{kl} plane showing the  location of \textit{Bmab} and \textit{P}2$_1$/\textit{a} fundamentals and both CDW and SDW superlattice reflections. \textbf{c} Crystal structure of Pr$_4$Ni$_3$O$_1$$_0$. The unit cell contains trilayer perovskite-like blocks that are related by an approximate B-centering operation
with the \textit{P}2$_1$/\textit{a} unit cells shown by black solid lines and
the related pseudo-orthorhombic unit cell by gray dashed lines. \textbf{d} - \textbf{e} Temperature evolution of the charge density wave order – (0 4.21 0.5). \textbf{d} Temperature dependence of an \textit{L} scan through the CDW peak with \textbf{e}, \textbf{f} showing the fitted amplitude and FWHM of this order, respectively.}
\label{Fig1}
\end{figure*}

\ch{RE4Ni3O10} (RE=Pr, La) is the $n=3$ member of the RP series and features three layers of corner-sharing \ch{NiO6} octahedra (Fig.~\ref{Fig1}a,c). Similar to some 214 cuprates, the \ch{NiO6} octahedra rotate relative to the \textit{c} axis by 10.9$^\circ$ for \ch{Pr4Ni3O10} and 6.84$^\circ$ for \ch{La4Ni3O10} towards the \textit{b} axis, leading to an in-plane Ni-O-Ni bond angle of 157.3$^\circ$ for \ch{Pr4Ni3O10} and 166.6$^\circ$ for \ch{La4Ni3O10}. These octahedral tilts give rise in both \ch{Pr4Ni3O10} and \ch{La4Ni3O10} to the monoclinic \textit{P}2$_1$/\textit{a} space group (Fig.~\ref{Fig1}c) \cite{zhang16, zhang2020high}. In terms of the electronic structure, the Ni site has an average valence of Ni$^{2.67}$ with a 3\textit{d}$^{7.33}$ configuration. Thus, analogous to commensurately doped cuprates \cite{tranquada1995evidence}, nickelates \cite{chen1993charge} and manganites \cite{mitchell2001spin}, the Ni valence would be expected to favor a site-by-site variation of the Ni valence, nominally Ni$^{3+}$-Ni$^{2+}$-Ni$^{3+}$, that would be stabilized by commensurate deformations of the underlying lattice.

Deviating from these expectations, intertwined CDW/SDW order exists in both \ch{La4Ni3O10} and \ch{Pr4Ni3O10} \cite{zhang2020intertwined, samarakoon2023bootstrapped}, with in-plane ordering wavevectors significaly different from the above expectation. For \ch{Pr4Ni3O10}, the in-plane SDW and CDW wavevectors are (0~0.61) and (0~0.78), respectively, well away from the anticipated value of (0~2/3) for both. The selection rules of the two order parameters are displayed in Fig.~\ref{Fig1}b. We adopt the unit cell for the monoclinic \textit{P}2$_1$/\textit{a} space group, while previous studies have used the pseudo-orthorhombic notation (\textit{Bmab}) to account for the twinning of crystallographic domains (the deviation of \textit{P}2$_1$/\textit{a} from \textit{Bmab} is small). Hereafter, for clarity, we will use the subscripts $\emph{o}$ and $\emph{m}$ to distinguish the Miller indices under the two notations, with the correspondence between the two notations given by $(H~K~L)_\emph{o} \equiv (H~K~L/2-H/2)_\emph{m}$. Unless otherwise stated, all the data and calculations in the main text concern \ch{Pr4Ni3O10}. Results for \ch{La4Ni3O10} can be found in the \textbf{Supplementary Information}.

In Fig.~\ref{Fig1}d, we present the temperature dependence of the elastic data at one of these CDW wavevectors $\bm{Q}_\text{CDW}$ = (0 4.21 0.5)$_\emph{m}$. The diffraction intensity was obtained using an IXS spectrometer with an energy resolution of 1.36 meV \cite{said2020high}. Further experimental details are presented in the \textbf{Methods}. The evolution of the diffraction intensity and the full-width-at-half-maximum (FHWM) of the CDW order with temperature are in Fig.~\ref{Fig1}e and Fig.~\ref{Fig1}f, respectively. The onset of the CDW is at $T_\text{CDW}\sim$ 157.5 K, consistent with previous findings \cite{zhang2020intertwined, samarakoon2023bootstrapped}. At the lowest temperature of 10 K in our measurement, the FWHM of the CDW order is 0.075 r.l.u. corresponding to a correlation length of approximately 6 nm along the \textit{c} axis. Fitting the temperature-dependent amplitude using a functional form of $(1-T/T_\text{CDW})^n$ yields a critical exponent $n$ $\sim$ 0.46 close to the mean-field value of 0.5, which indicates a second-order phase transition. In previous studies \cite{zhang2020intertwined, samarakoon2023bootstrapped}, the critical exponent for the SDW order is essentially identical to that of the CDW, which would suggest that both the CDW and SDW are primary order parameters. In comparison, in commensurately-doped cuprates \cite{tranquada1995evidence}, the critical exponent of the charge order is smaller than that of the spin order. Beyond strongly correlated oxides, for chromium, an established CDW/SDW material, the CDW order is secondary with a critical exponent twice that of the SDW \cite{fawcettRMP}. Thus, examining the charge order in more detail should provide crucial insights into the nature of the intertwined CDW/SDW and the potential role of electron-phonon coupling in the trilayer nickelates.

\subsection{Persistent phonon dispersions near the CDW wavevector}

\begin{figure*}
\includegraphics[width=6. in]{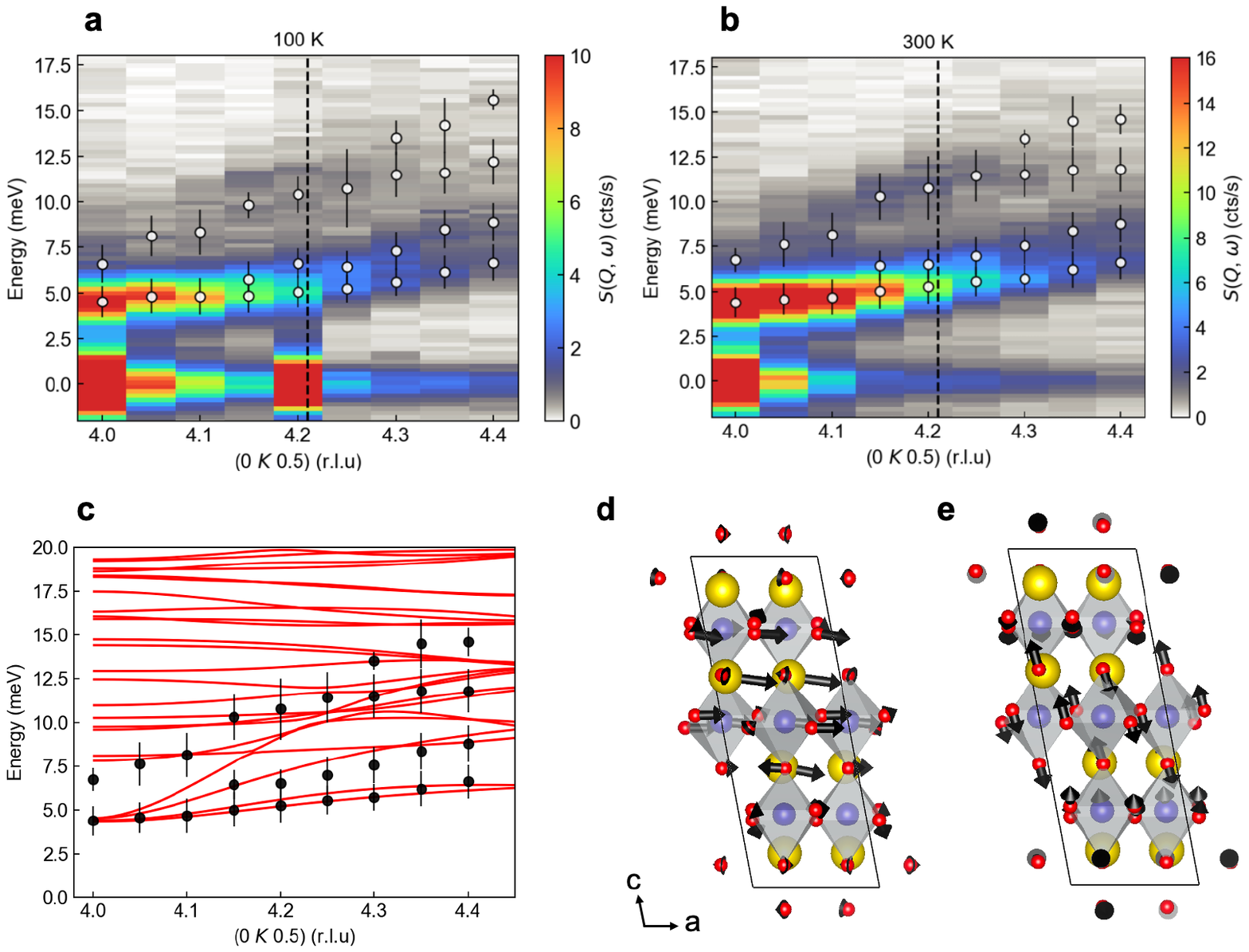} %
\caption{\textbf{Phonon dispersions below and above $T_{CDW}$ and the eigenvectors of the phonon modes near the charge density wave order – (0 4.21 0.5).} \textbf{a} and \textbf{b} 2D map of the phonon dispersions at 100 K and 300 K, respectively. The black empty circles correspond to the energy of the phonon modes and the vertical bars to their peak widths. The vertical dashed line marks the CDW wavevector (0 4.21 0.5). \textbf{c} calculated phonon dispersions (solid red curves) with the fitting results from experiments (solid black circles with error bars). \textbf{d} and \textbf{e} Eigenvectors of the phonons with energies of 5.28 meV and 7.31 meV at the CDW wavevector – (0 4.21 0.5).}
\label{Fig2}
\end{figure*}

We performed non-resonant IXS to understand the electron-phonon coupling in \ch{RE4Ni3O10}, focusing on the lower energy phonon branches that are accessible by this technique. Fig.~\ref{Fig2}a and \ref{Fig2}b show the measured phonon spectra between (0~4~0.5)$_\emph{m}$ and (0~4.5~0.5)$_\emph{m}$ at 100 K (below $T_\text{CDW}$) and at 300 K (above $T_\text{CDW}$), respectively. Up to four phonon branches are distinguishable and are marked using white solid circles.

The CDW ordering wavevector $\bm{Q}_\text{CDW}$ is marked using a vertical dashed line, with pronounced elastic scattering intensity at 100 K, demonstrating the presence of charge order. The temperature dependence of the elastic scattering from the CDW order is presented in Fig.~\ref{Fig1}c-e as noted above. The measured phonon dispersions and spectral weights do not change across the phase transition within our experimental resolution, with no discernible phonon softening at $\bm{Q}_\text{CDW}$. Notably, in monoclinic notation, the CDW wavevector has an out-of-plane component of $L=0.5$. As such, $\bm{Q}_\text{CDW}$ is away from any high symmetry direction of the Brillouin zone. As a result, at (0~4~0.5)$_\emph{m}$, the observed phonon energies remain finite. More importantly, the eigenvectors corresponding to the observed phonon modes cannot be simply characterized as longitudinal or transverse. This actually reduces the possibility that the lattice deformation responsible for the CDW is completely orthogonal to the phonon modes we captured and hence escaped detection.

To delve into this further, we calculated the phonon dispersions using the phonon density functional theory (PDFT) (Fig.~\ref{Fig2}c). Details of the PDFT are in the \textbf{Methods}. The calculated phonon dispersions agree to the most part with the experimental results, especially near (0~4~0.5)$_\emph{m}$. There is a larger discrepancy for the highest phonon branch near (0~4.5~0.5)$_\emph{m}$. This most likely arises from the large number of phonon branches due to the low monoclinic symmetry and an energy-varying phonon spectral weight in the IXS measurements.

The absence of a Kohn anomaly in any of these phonon branches is striking. In comparison, a phonon softening of 2 meV is seen in La$_{1.875}$Ba$_{0.125}$CuO$_4$ \cite{miao2018incommensurate} and as high as 5 meV in YBa$_2$Cu$_3$O$_{6.6}$ \cite{le2014inelastic} for phonons in a similar energy range, though only one of the two detected branches softened in \cite{miao2018incommensurate} while both softened in \cite{le2014inelastic}. The softened phonon branches should, in principle, have atomic displacements relevant to the formation of the CDW. To this end, we calculated the eigenvectors of the measured phonon branches, encompassing ionic motions both in and out of the Ni-O planes. In Figs.~\ref{Fig2}d and \ref{Fig2}e, we show the calculated eigenvectors in the monoclinic unit cell for two of these phonon branches. For the lowest-lying branch, the Ni and O atoms predominantly vibrate along the \emph{a} axis and with a significant shearing between the center Ni-O plane and the outer ones (Fig.~\ref{Fig2}d). The higher-energy mode corresponds to a superposition of the breathing of the apical oxygens and an octahedral rotation relative to the Ni-O planes. 
There are also large A-site displacements for the eigenvectors of all four phonon branches we could observe, and none along any high symmetry direction. This reduces the possibility that the phonon softening was not observed for symmetry reasons. In other correlated oxides, these deformations of the metal-oxygen octahedra help stabilize charge order. Therefore, since the four phonon modes we could resolve should include vibrations affected by the CDW order, the lack of any softening is striking.

\begin{figure*}
\includegraphics[width=5 in]{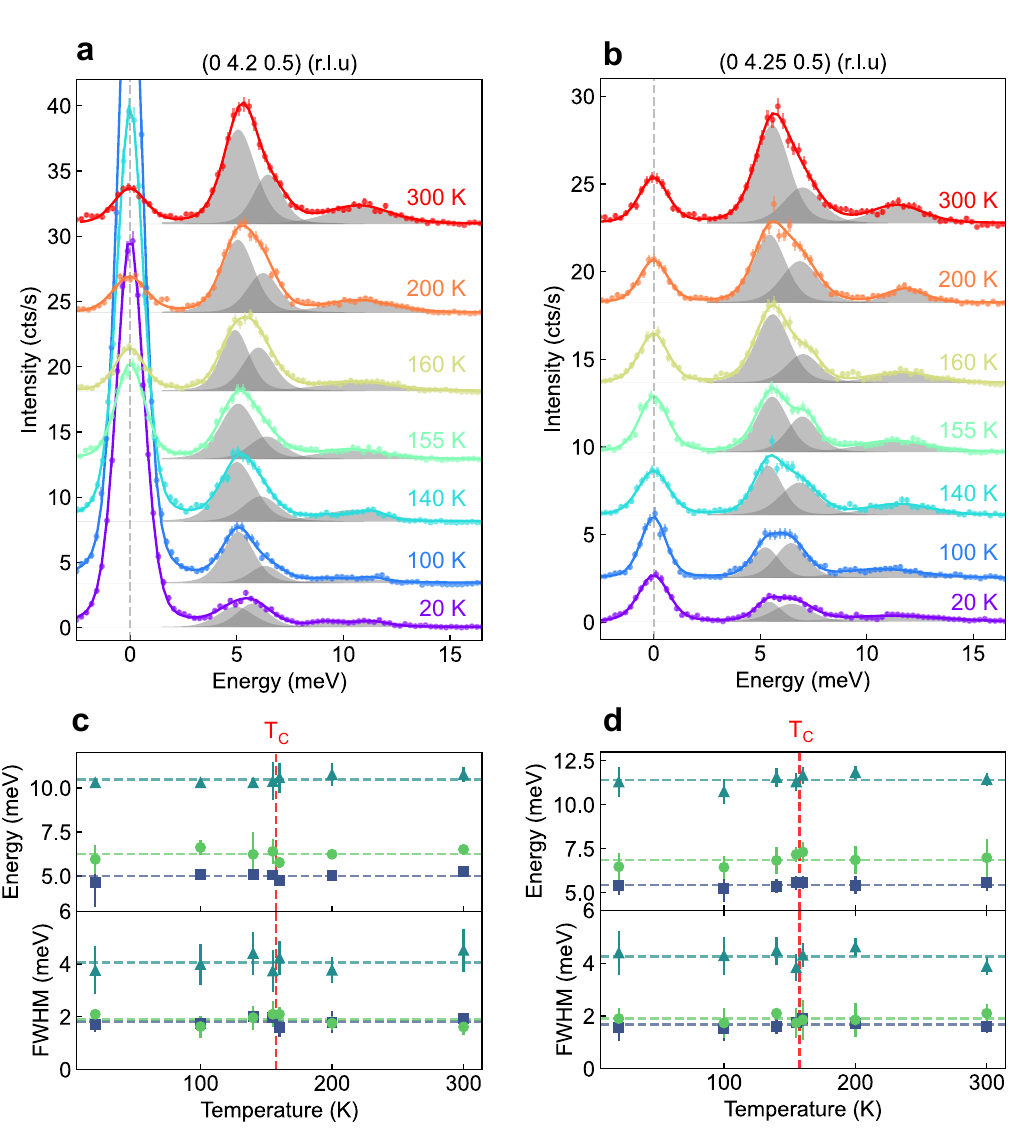} %
\caption{\textbf{Temperature-dependent phonon line shape analysis around the CDW wavevector – (0 4.21 0.5).} \textbf{a} and \textbf{b}. Temperature dependence of the constant-$\bm{Q}$ phonon measurements at $\bm{Q}$ = (0 4.20 0.5) and (0 4.25 0.5), respectively. The phonons, fitted as shown by the gray shading, are observed by meV-resolved inelastic X-ray scattering. \textbf{c} and \textbf{d} Temperature dependence of the fitted energy positions and FWHMs of the phonons at $\bm{Q}$ = (0 4.20 0.5) and (0 4.25 0.5), respectively. The red dashed lines mark $T_{CDW}$.}
\label{Fig3}
\end{figure*}

To rule out phonon softening at $Q_\text{CDW}$, we track the detailed temperature dependence of the phonon dispersions. Figs.~\ref{Fig3}a and \ref{Fig3}b depict the IXS spectra at (0~4.2~0.5)$_\emph{m}$ and (0~4.25~0.5)$_\emph{m}$. Among the two, (0~4.2~0.5)$_\emph{m}$ is essentially at $\bm{Q}_\text{CDW}$, with a separation well within the FWHM of the CDW wavevector. The elastic spectral weight increases below T$_\text{CDW}$ following the onset of the order parameter. Fitting procedures identical to those in Fig.~\ref{Fig2} were used, with the contribution from each branch shown as shaded areas underneath the corresponding phonon spectra. For the two momentum transfers shown in Fig.~\ref{Fig3}, consistent with Fig.~\ref{Fig2}, three phonon branches were used to optimize the fit.

The energy and the FWHM of the three phonon branches are displaced in Figs.~\ref{Fig3}c and \ref{Fig3}d for (0~4.2~0.5)$_\emph{m}$ and (0~4.25~0.5)$_\emph{m}$ respectively, which are essentially unchanged. Indeed, no change in the phonon dispersions were observed along the entire (0, K, 0.5)$_\emph{m}$ cut. Further discussion about the temperature evolution can be found in the \textbf{Supplementary Information}.

\subsection{The role of SDW on the emergence of the intertwined order}


The absence of any phonon anomalies at or near $\bm{Q}_{CDW}$ is highly nontrivial. It begs the question of whether any phonon softening occurs in the material, and if it does, where in energy and in reciprocal space this would occur and how much softening would be present. 
For instance, large phonon anomalies were seen in optical branches at much higher energies in both cuprates and 214 nickelates \cite{tranquada2002bond, merritt2020giant}.  Unfortunately, our IXS setup does not allow enough scattered X-ray photons to study these high-energy phonons. Nonetheless, as demonstrated earlier in the paper, our observation is dramatically different from other correlated oxides where the robust charge order is associated with insulating-like behavior. Instead, in the 4310 nickelates, the intertwined order is associated with a metal-metal transition, similar to what is observed in chromium.

\begin{figure*}
\includegraphics[width=6. in]{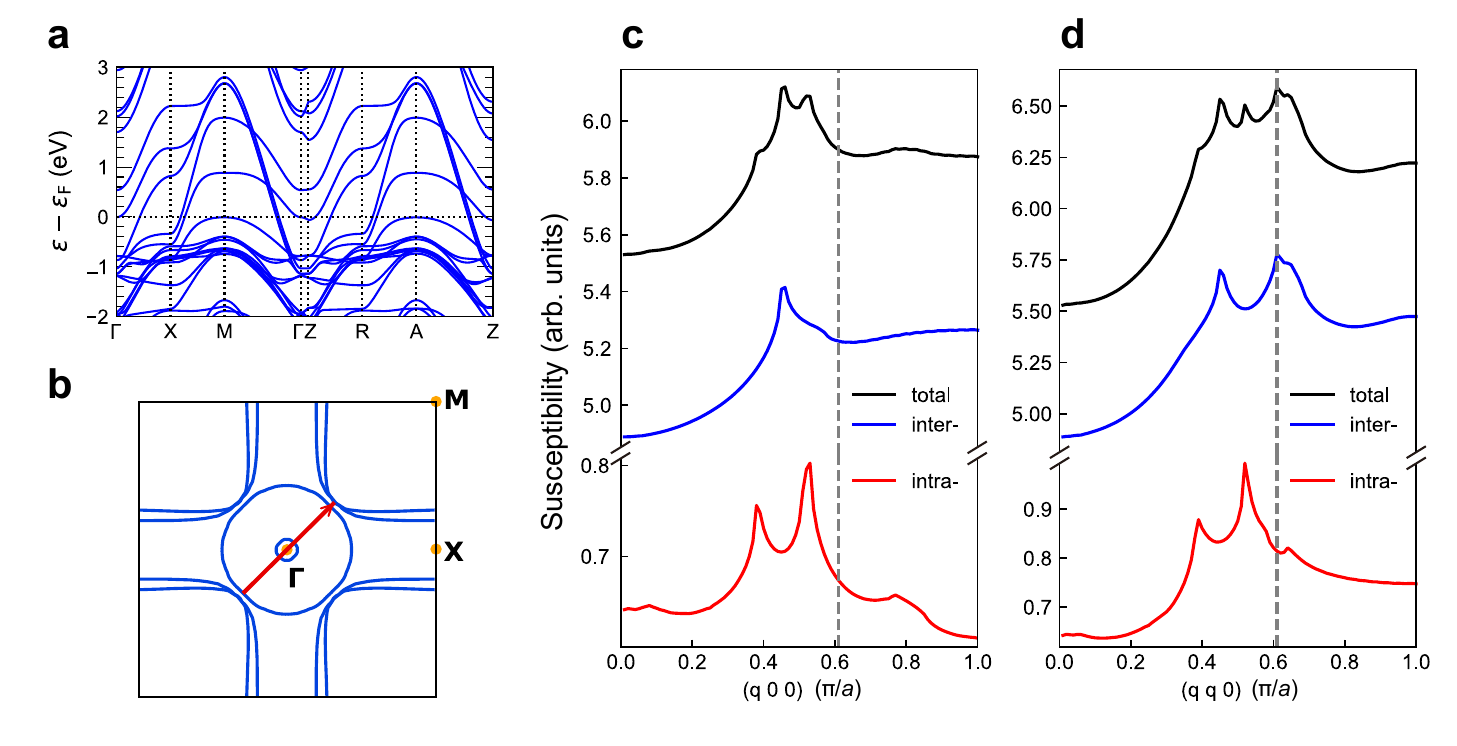} %
\caption{\textbf{The origin of the SDW order in Pr$_4$Ni$_3$O$_1$$_0$.} \textbf{a} Band structure of Pr$_4$Ni$_3$O$_1$$_0$ along high symmetry cuts in the tetragonal \textit{I}4/\textit{mmm} zone. \textbf{b} Fermi surface contours of Pr$_4$Ni$_3$O$_1$$_0$ with the nesting vector indicated in red. \textbf{c} and \textbf{d} Susceptibility showing the total contribution, interband contribution, and intraband contribution along (q 0 0) and (q q 0), respectively, with the observed SDW wavevector indicated by the vertical dashed line in \textbf{d}. The predicted order occurs along the zone diagonal at the observed SDW wavevector (0.61 0.61 0) in tetragonal notation (the $L$ dependence, due to magnetic coupling between layers, is not captured by $\chi_0$).  No structure is seen at the observed CDW wavevector.}
\label{Fig4}
\end{figure*}

Therefore, a more likely scenario has to do with the unique electronic structure of the trilayer nickelates. To demonstrate the relation of the CDW/SDW to the electronic structure, we examine the potential role of Fermi surface nesting by calculating the bare susceptibility function
\begin{eqnarray}
    \chi_0(\bm{Q}) = 2 \sum_{n,n^\prime}\int d\bm{k}\frac{f_{\bm{k}+\bm{Q},n^\prime}-f_{\bm{k},n}}{\epsilon_{\bm{k}+\bm{Q},n^\prime}-\epsilon_{\bm{k},n}}\label{eqn:lindhard}
\end{eqnarray}
with the full $\chi$ given by $\chi_0/(1- I \chi_0(\bm{Q}))$ where $I$ is the interaction associated with the order (the denominator of $\chi$ going to zero marks the onset of mean-field order). Here $f$ is the Fermi function and $\epsilon$ are the band energies with $n$,$n^\prime$ band indices. The factor of 2 takes into account the sum over spin. For \ch{La4Ni3O10}, the band structure and susceptibility were calculated previously using the orthorhombic $Bmab$ structure \cite{zhang2020intertwined}. One issue with that calculation was the neglect of matrix elements in Eq.~\ref{eqn:lindhard} (that is, of the band wavefunctions with the change/spin operators). This effect can be mitigated by using an unfolded Brillouin zone~\cite{gupta}. Therefore, we calculated the band structure and the corresponding Fermi surface of \ch{Pr4Ni3O10} using the tetragonal $I4/mmm$ unfolded structure  (see Figs.~\ref{Fig4}a-b). As shown in Fig.~\ref{Fig4}b, the large electron pocket centered around $\Gamma$ can be connected to the two large hole pockets centered around $M$ by a vector equal to the observed SDW ordering vector, while we observe no obvious nesting at the observed CDW vector. This is confirmed by calculations of $\chi_0(\bm{Q})$ (Figs.~\ref{Fig4}c-d) where a peak in $\chi_0$ is obtained at the observed SDW vector, which can be attributed to the interband terms referred to above, with no peak observed at the CDW wavevector.

The presence of a susceptibility peak at $\bm{Q}_{SDW}$ demonstrates that spin rather than charge is the driving force behind the onset of the intertwined order. This calculation is consistent with the absence of a Kohn anomaly in our data at the CDW wavevector. In principle, phonon softening could be present at the SDW wavevector as observed in chromium \cite{lamago2010measurement}. However, IXS data we have taken at one of the SDW ordering vectors also found no evidence for phonon softening. Such an observation further establishes the critical role of the spin degree of freedom. We further comment on the temperature dependence of the SDW and CDW order parameters (Fig.~\ref{Fig1}). That both order parameters share similar critical exponents may be a result of the intertwined order being proximate to a first-order transition. Details can be found in \cite{norman2025landau}.

The temperature independence of the phonon dispersions is unusual not only among correlated oxides, but even compared to CDW and SDW materials with weaker electronic correlations. For example, the CDW in transition metal dichalcogenides such as 2H-\ch{NbSe2} were initially perceived as driven by the electronic structure. However, in the majority of these dichalcogenides, phonons soften and become overdamped substantially when cooling down towards the CDW phase transition, e.g., by $\sim$8 meV in 2H-\ch{NbSe2} \cite{weber2011extended}, followed by the expected hardening below the phase transition. Similarly for Cr, phonons soften as a function of temperature not only at the SDW wavevector, but over extended regions of the zone boundary in reciprocal space \cite{lamago2010measurement}. In this sense, the intertwined order in \ch{RE4Ni3O10} presents a more-perfect electronic order closer to theoretical proposals, which could provide a much-sought-after playground for understanding the emergent properties of such states.

\section{III. Conclusion}

The unaltered-with-temperature phonon dispersions in the trilayer nickelates that we observed are markedly different from those in other correlated oxides and potentially represent a new regime of intertwined order stabilized by the strong interaction of the spin degree of freedom and associated with a metal-metal transition. Fully elucidating the role of the lattice in the formation of the CDW/SDW requires further experimental and theoretical efforts, including a three-dimensional refinement of any potential lattice distortions associated with the phase transition, as well as calculations of the electron-phonon coupling across the entire Brillouin zone. Nonetheless, the new regime we discovered provides a valuable playground both for understanding the nature of the electronically intertwined order weakly coupled to the lattice. Since non-conventional superconductivity has been reported in trilayer RP nickelates, our findings lay the groundwork for revealing the cooperative vs. competing relationship between the intertwined order and superconductivity in and beyond nickelates.

\section{IV. Methods}

\subsection{A. Material synthesis and sample preparation} 

\ch{Pr4Ni3O10} crystals were grown in a high-pressure optical floating zone furnace (HKZ 150-bar Model, ScIDre, Germany) \cite{zhang2017large,zhang2020high}. Stoichiometric \ch{Pr6O11} and NiO powders were thoroughly mixed and sintered at 1050$^\circ$C for 36 h with several inter-grindings to obtain the precursor powder of \ch{Pr4Ni3O10}. Precursor powder was pressed into a rod with a diameter of $\sim$ 6 mm by hydrostatic press and then sintered for 24 h at 1050$^\circ$C. \ch{Pr4Ni3O10} crystals were grown via a two-step method at an oxygen pressure of 140 bar using a 5 kW xenon lamp as the heating source: The first step was a fast pass and melt with a traveling rate of 30-50 mm/h to obtain a rod with improved density, and the second step was the slow growth using the densified rod with a traveling rate of 4-5 mm/h. During the growth, the feed and seed rods were counter-rotated at 15-25 rpm, and an oxygen flow rate of 0.1 L/min was maintained. Single crystals were mechanically separated from the as-grown boule.

Single crystals were aligned using a lab-based four-circle X-ray diffractometer with a photon energy of 8.05 keV at room temperature to confirm the crystal is in pure \textit{P}2$_1$/\textit{a} phase. To meet the experimental requirements of inelastic X-ray scattering and have good statistics from inelastic scattering signals, the crystal was further polished with the thickness down to around 150 $\mu$m. 

\subsection{B. Inelastic X-ray Scattering} 

The experiment was performed at beamline 30-ID-C (HERIX) of the Advanced Photon Source (APS). The highly monochromatic X-ray beam with an incident energy of 23.7 keV was focused on the sample with a cross section of $\sim$ 35 $\times$ 15 $\mu$m\textsuperscript{2} (horizontal $\times$ vertical). The total energy resolution of the monochromatic X-ray beam and analyzer crystals was $\Delta E\sim$ 1.36 meV (full width at half maximum (FWHM)). The phonon measurements were performed in the Laue geometry, while the diffraction was performed in transmission geometry. Typical counting times ranged from 30 sec to 1 min per point in the energy scans at constant momentum transfer \textit{Q}.

\subsection{C. Phonon density functional theory (PDFT) calculations}

The phonon spectrum of Pr$_4$Ni$_3$O$_{10}$ ($P2_{1}/a$) is computed from first-principles using the frozen phonon method as implemented in Phonopy interfaced with VASP~\cite{KresseVASP1,KresseVASP2}. We employed the Perdew-Burke-Ernzerhof (PBE) exchange-correlation functional~\cite{gga_pbe}, a plane-wave energy cutoff of 500 eV, and a Gamma-centered $2\times8\times8$ $k$-mesh with 0.1 eV Gaussian smearing.

\subsection{D. Susceptibility calculations}

Eq.~\ref{eqn:lindhard} is written in the approximation that matrix elements between the band wavefunctions and the charge/spin operators are not accounted for. In such an approximation, it is best to use an unfolded Brillouin zone~\cite{gupta}. Hence, we compute the band structure of Pr$_{4}$Ni$_{3}$O$_{10}$ for the high-symmetry tetragonal ($I4/mmm$) structure using the all-electron, full potential DFT code Wien2k~\cite{Blaha2020wien2k} with the PBE exchange-correlation functional. The basis set size was set by RK$_{\mathrm{max}}$ = 7 and R$_{\mathrm{MT}}$ = 2.37, 1.90, and 1.69 for Pr, Ni, and O, respectively. Brillouin zone integrations were done on a $21\times21\times21$ $k$-mesh. The Pr($4f$) electrons are treated as core electrons within the open-core approximation.

To calculate the static bare susceptibility $\chi_0(\vec{Q})$, we interpolate the five bands nearest the Fermi energy using a Fourier spline series~\cite{koelling}, with 2436 functions fit to 726 $k$-points in the irreducible wedge of the Brillouin zone. Eq.~\ref{eqn:lindhard} was then evaluated using a tetrahedron decomposition of the Brillouin zone ($3 \times 8^5$ tetrahedra in the irreducible wedge)~\cite{rath}.

\section{Acknowledgements} The sample synthesis, IXS measurements, data analysis, and susceptibility calculations at Argonne National Laboratory were supported by the U.S. Department of Energy, Office of Science, Basic Energy Sciences, Materials Science and Engineering Division. Work at Brookhaven (X-ray scattering and data interpretation by Y.S.\ and M.P.M.D.) was supported by the U.S.\ Department of Energy, Office of Science, Office of Basic Energy Sciences, under Award No.~DE-SC0022311. A.S.B. and H.L. acknowledge support from NSF Grant No. DMR-2323971. This research used resources of the Advanced Photon Source, a U.S. Department of Energy (DOE) Office of Science User Facility operated for the DOE Office of Science by Argonne National Laboratory under Contract No. DE-AC02-06CH11357.


\section{Data availability} The data and the analysis codes will be available upon reasonable request.


%

\end{document}